\documentclass[]{aipproc}

\layoutstyle{6x9}

\SetInternalRegister\hbadness{8000} 

\usepackage{epsfig}
\usepackage[scriptsize]{subfigure} 
\usepackage{amsmath}
\usepackage{amssymb}
\usepackage{graphics}

\newcommand{\pppKL}{\mbox{$pp \rightarrow pK^{+}\Lambda$}\ }
\newcommand{\pppKSo}{\mbox{$pp \rightarrow pK^{+}\Sigma^{0}$}\ }

\newcommand{\be}{\begin{equation}}
\newcommand{\ee}{\end{equation}}
\newcommand{\bea}{\begin{eqnarray}}
\newcommand{\eea}{\end{eqnarray}}

\begin{document}

\title[Energy dependence of the $\Lambda / \Sigma^0$  production cross section
	ratio]
	{Energy dependence of the $\Lambda / \Sigma^0$  production cross section
	ratio in p--p interactions.}


\newcommand{\ikpjuel}{IKP, Forschungszentrum J\"{u}lich, D-52425 J\"{u}lich, Germany}
\newcommand{\ikpmue}{IKP, Westf\"{a}lische Wilhelms--Universit\"{a}t, D-48149 M\"{u}nster, Germany}
\newcommand{\cracow}{M.~Smoluchowski Institute of Physics, Jagellonian University, PL-30-059 Cracow, Poland}
\newcommand{\nphycracow}{H. Niewodnicza{\'n}ski Institute of Nuclear Physics, PL-31-342 Cracow, Poland}
\newcommand{\catowice}{Institute of Physics, University of Silesia, PL-40-007 Katowice, Poland}
\newcommand{\zeljuel}{ZEL,  Forschungszentrum J\"{u}lich, D-52425 J\"{u}lich,  Germany}

\author{P.~Kowina}{address={\ikpjuel}}
\author{H.-H.~Adam}{address={\ikpmue}}
\author{A.~Budzanowski}{address={\nphycracow}}
\author{R.~Czy{\.{z}}ykiewicz}{address={\cracow,\ikpjuel}}
\author{D.~Grzonka}{address={\ikpjuel}}
\author{M.~Janusz}{address={\cracow}}
\author{L.~Jarczyk}{address={\cracow}}
\author{B.~Kamys}{address={\cracow}}
\author{A.~Khoukaz}{address={\ikpmue}}
\author{K.~Kilian}{address={\ikpjuel}}
\author{P.~Moskal}{address={\ikpjuel}}
\author{W.~Oelert}{address={\ikpjuel}}
\author{C.~Piskor-Ignatowicz}{address={\cracow}}
\author{J.~Przerwa}{address={\cracow}}
\author{T.~Ro{\.{z}}ek}{address={\ikpjuel}}
\author{R.~Santo}{address={\ikpmue}}
\author{G.~Schepers}{address={\ikpjuel}}
\author{T.~Sefzick}{address={\ikpjuel}}
\author{M.~Siemaszko}{address={\catowice}}
\author{J.~Smyrski}{address={\cracow}}
\author{A.~Strza{\l}kowski}{address={\cracow}}
\author{A.~T\"aschner}{address={\ikpmue}}
\author{P.~Winter}{address={\ikpjuel}}
\author{M.~Wolke}{address={\ikpjuel}}
\author{P.~W{\"u}stner}{address={\zeljuel}}
\author{W.~Zipper}{address={\catowice}}
 \copyrightholder{Acoustical Scociety of America}
\copyrightyear  {2001}

\begin{abstract}
Measurements of the near threshold $\Lambda$ and $\Sigma^0$ production
via the $pp \rightarrow p K^+ \Lambda / \Sigma^0$ reaction at COSY--11 
have shown that the $\Lambda/\Sigma^0$ cross section ratio exceeds the 
value at high excess energies ($\mbox{Q}~\ge$~300~MeV) by an order of 
magnitude. For a better understanding additional data have been taken
between 13~MeV and 60~MeV excess energy.  \\
Within the first $20~\mbox{MeV}$ excess energy a strong decrease of the cross
section ratio is observed, with a less steep decrease in the higher excess
energy range.\\
A description of the data with a parametrisation including $p-Y$ 
final state interactions suggests a much smaller $p-\Sigma^0$ FSI 
compared to the $p-\Lambda$ system.
\end{abstract}

\date{\today}

\maketitle

\section{Introduction}
	At the COSY--11 facility~\cite{bra96} measurements of the 
$\Lambda$ and $\Sigma^0$ hyperon production were performed via the \pppKL and 
\pppKSo reactions close to threshold~\cite{sew99} resulting in a cross 
section $\sigma (\Lambda)$ for the $\Lambda$ production which is more then one
order of magnitude larger than the cross section $\sigma (\Sigma^0)$ for the 
$\Sigma^0$ production.

	Since the quark contents of these two hyperons are the same,
based on the isospin relations only, the ratio of the cross sections 
${\cal R} = \sigma (\Lambda) / \sigma ( \Sigma^0)$ should be equal to three.
In fact at high excess energies~\cite{bal88} a ratio of $\sim2.5$ is observed 
in contrast to the by more then one order of magnitude larger ${\cal R}$ at 
threshold.
	
	In order to understand this behavior, measurements~\cite{kow03} in the 
intermediate energy range ($13~\mbox{MeV} \le Q \le 60~\mbox{MeV}$), where
the ratio ${\cal R}$ was expected to decrease from $\sim 28$ to $\sim 2.5$, 
have been performed.

\section{Experiment}
\label{experiment}

\begin{figure}[hbt]
\epsfig{file=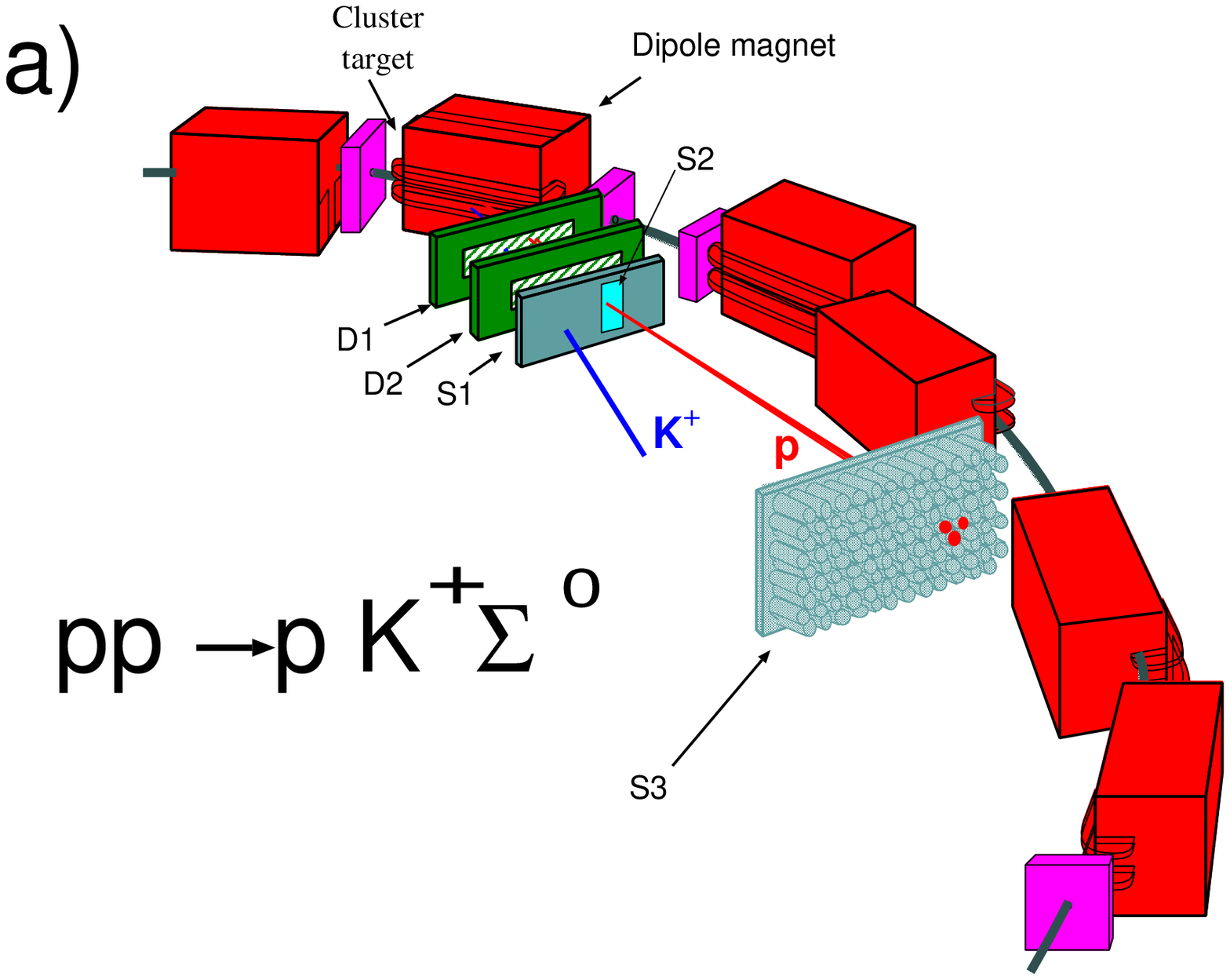,scale=0.3}
\hspace{1cm}
\epsfig{file=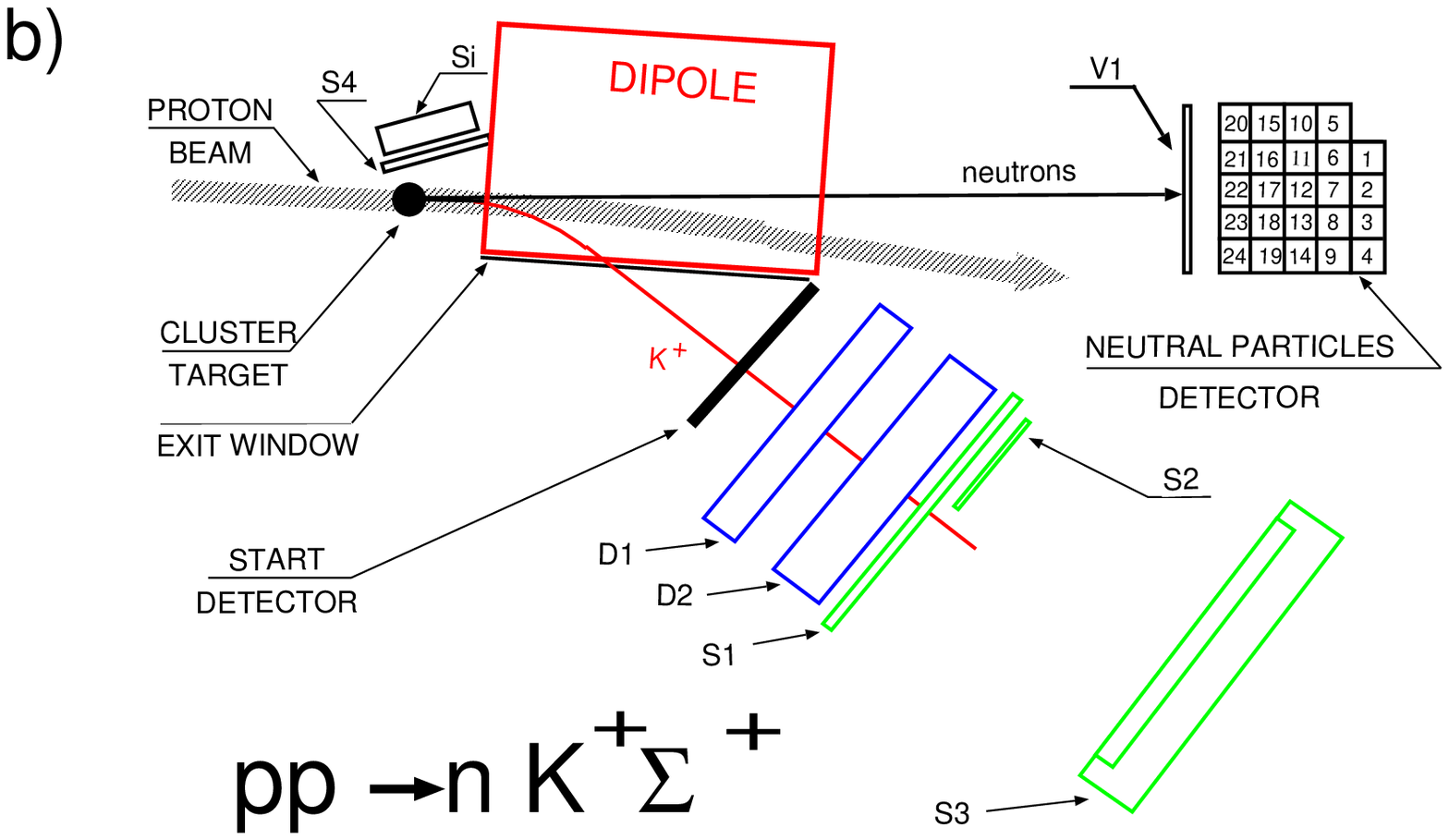,scale=0.4}
\caption{a) COSY--11 facility. \:b) Setup extended by the start and neutral 
particle detectors used in the measurements of the $pp \to nK^+\Sigma^+$ 
reaction.}
\label{cosy11}
\end{figure}

	The measurements of the hyperon production were performed at the 
COSY--11 facility~\cite{bra96,pawel:nim} (see figure~\ref{cosy11}) at the 
{\bf{Co}}oler {\bf{Sy}}n\-chrotron COSY-J\"ulich~\cite{maie97}.

	One of the regular COSY dipole magnets serves as a magnetic 
spectrometer with a H$_{2}$ cluster beam target~\cite{domb97} installed in 
front of it. The interaction between a proton of the beam with a proton of the
cluster target may lead to the production of neutral hyperons 
($\Sigma^0$, $\Lambda$) via the reactions $pp \to pK^+\Lambda(\Sigma^0)$.

	Events of the $pK^+\Lambda(\Sigma^0)$ production are selected
by the detection of both positively charged particles in the exit channel 
(i.e.\ proton and $K^{+}$).  The unobserved neutral particle is identified via 
the missing mass method.

Positively charged ejectiles  are directed from the circulating beam by the 
magnetic field of the dipole towards the inner part of the COSY ring, where 
they are registered in a set of two drift chambers D1 and D2 for the track 
determination. Their momenta are reconstructed by tracking back the particles 
through the well known magnetic field to the assumed interaction point. The 
velocities of the ejectiles are given by a measurement of the time of flight
between the S1(S2) start and the S3 stop scintillator hodoscopes from which in 
combination with the momentum the invariant mass of the particle is given.
Therefore, the four-momentum vectors for all positively charged particles  are
known and the four-momentum of the unobserved neutral hyperon is uniquely 
determined.

	To avoid systematical uncertainties as much as possible, COSY was 
operated in the ''supercycle mode'' i.e.\ the beam momenta were changed 
between the cycles, such that for example 10 cycles with a beam momentum 
corresponding to the excess energy $Q=20$~MeV above the $\Sigma^0$ threshold 
were followed by one cycle with the same $Q$ above the $\Lambda$ production 
threshold. The ratio of the number of the cycles was chosen inversely 
proportional to the ratio of the expected cross sections for the $\Lambda$ and 
$\Sigma^0$ production. Thus, both cross sections were measured under the same 
conditions and possible changes in the detection system did not influence the 
data taking procedure, especially for the determination of the cross section 
ratio.

	The extention of the detection system by an additional neutral 
particle detector (see figure~\ref{cosy11}b) allows for the measurements of  
neutrons in the exit channel. This upgrade of the detection system allows
to extend the study reported in this paper into the production of charged
hyperons, e.g \ via the $pp \to nK^+ \Sigma^+$ reaction. To increase the 
acceptance an additional start detector for $K^+$ was installed in the system.

\section{Results}
\begin{figure}[hbt]
\vspace{-0.2cm}
\epsfig{file= 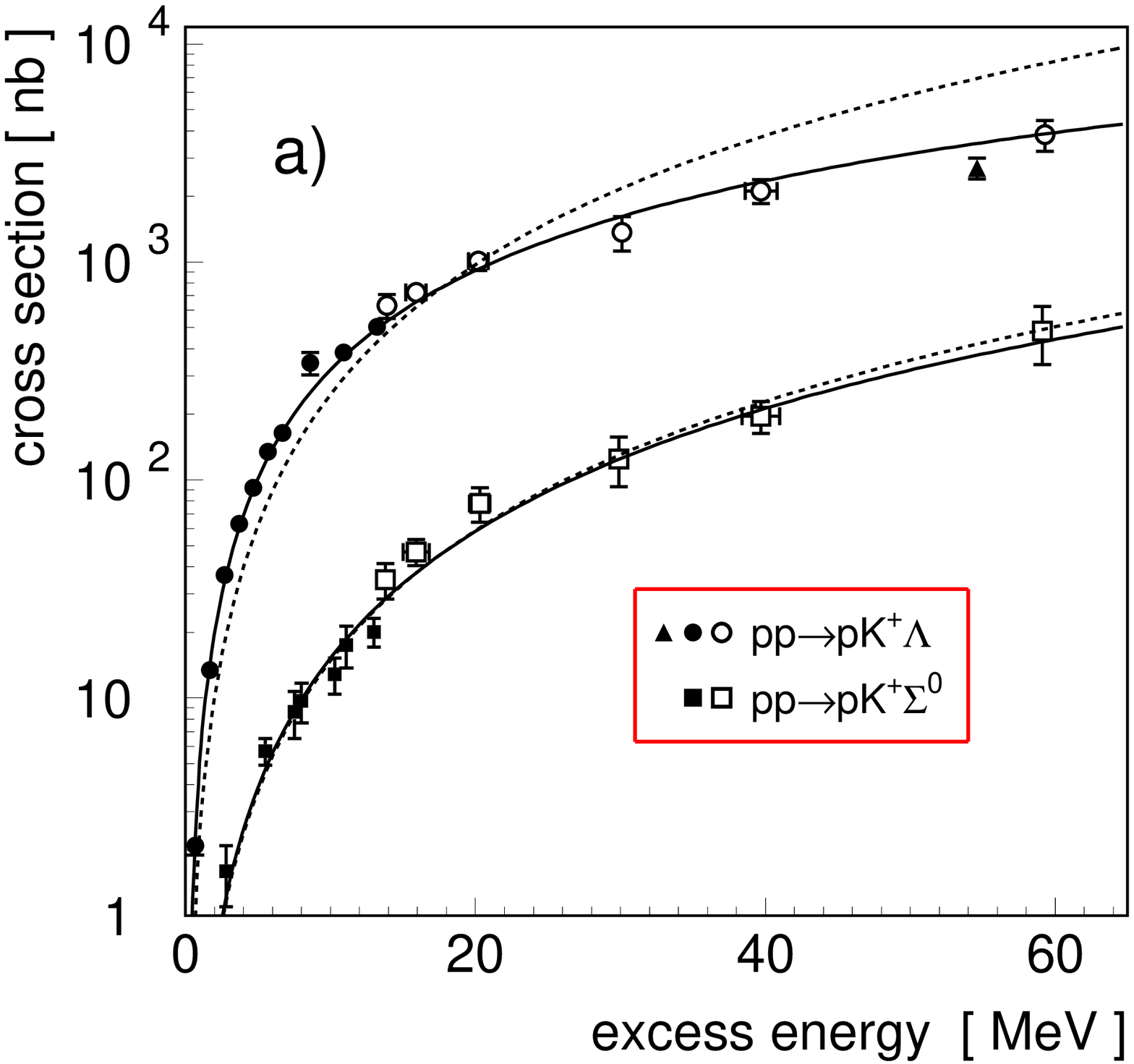,scale=0.3}
\epsfig{file= 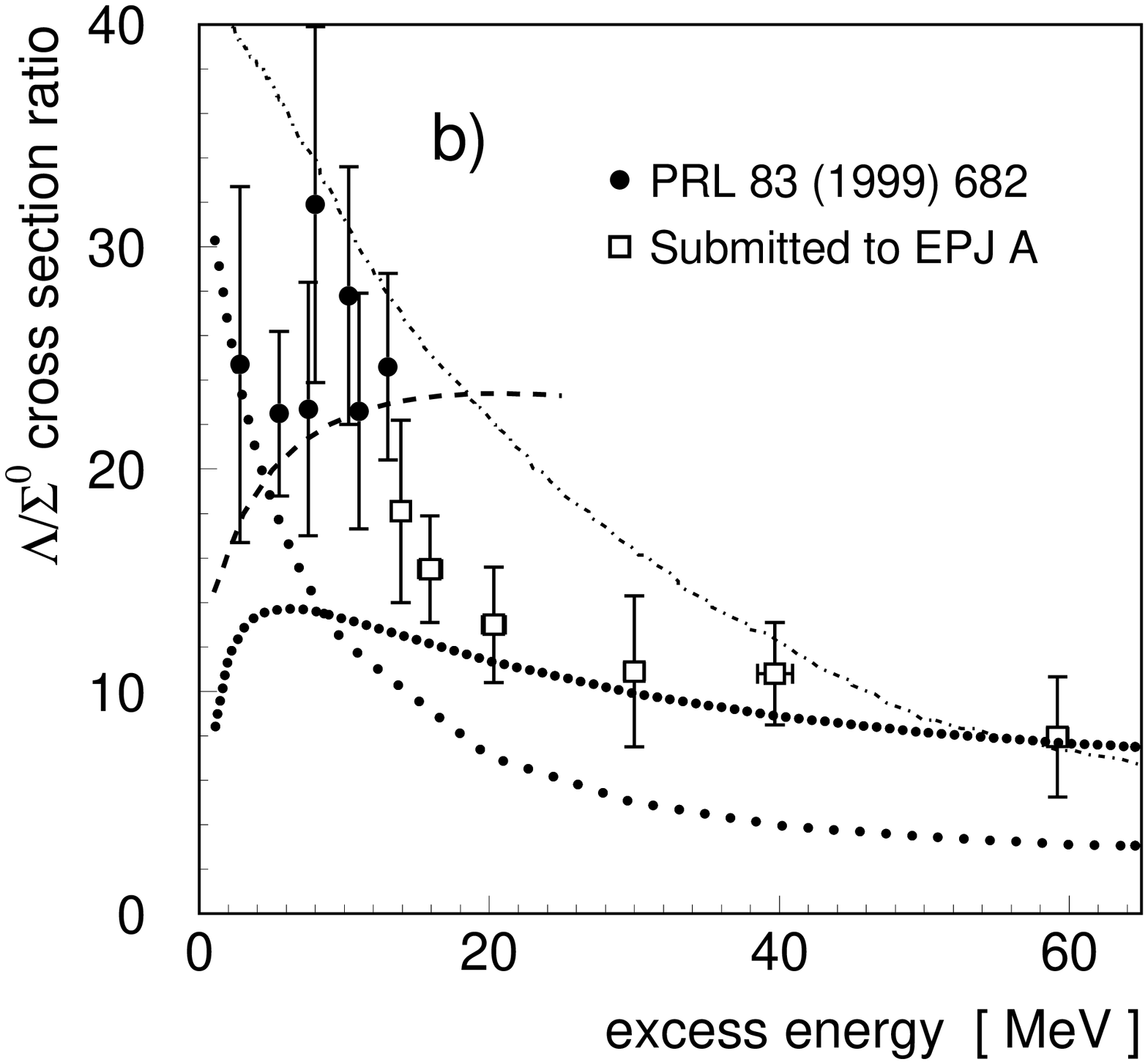,scale=0.3,angle=0.}
\caption{a) Total cross sections for the \pppKL and \pppKSo production
(full symbols~\cite{sew99,bal98a}, open symbols~\cite{kow03} and triangle~\cite{tof:blig98}).
\: b) Cross section ratio  for the $pp \to pK^+ \Lambda$ and $pp \to pK^+ \Sigma^0$ reactions. 
}
\label{cros:plot}
\end{figure}

	The hyperon production via $pp \to pK^+ \Lambda (\Sigma^0)$ has been 
studied in the excess energy range between $13$ and $60~\mbox{MeV}$. In 
figure~\ref{cros:plot}a) and figure~\ref{cros:plot}b) the excitation functions 
and the energy dependence of the cross section ratio are shown, respectively.
The most drastic decrease of the cross section ratio is observed between 
$10~\mbox{MeV}$ and $20~\mbox{MeV}$ following by a less steep decrease towards 
higher $Q$-values. 
	
	The first published close--to--threshold  data~\cite{sew99} have 
triggered many theoretical discussions. The results of available calculations 
are shown in figure~\ref{cros:plot}b) and are briefly discussed in the 
following section.

\vspace{-.6cm}
\subsection{Comparison with theoretical predictions} \label{comparision}

	Presently different theoretical calculations with various dominant 
production mechanisms are available which reproduce at least the trend of the 
data, see figure~\ref{cros:plot}b).

	Calculations by Sibirtsev, Tsushima et al.~\cite{sib00b,tsu99} were 
performed within two different models. In the first one -- Boson Exchange Model
-- (dense dotted line) pion and Kaon exchange is considered as the most 
important mechanism of the hyperon production~\cite{sib95}. The second model
(dotted line) bases on the assumption that the hyperon is produced in the 
decay of $N^*$ resonances excited via the exchange of $\pi$, $\eta$ and $\rho$
 mesons~\cite{tsu99,tsu97}. 

	Hyperon production via $N^*$ resonances was also investigated by Shyam 
et al.~\cite{shy01} (dashed-dotted) where $N^*(1650)$, $N^*(1710)$, $N^*(1720)$
 were assumed to be excited by the exchange of the $\pi$, $\rho$, $\omega$ and
$\sigma$ mesons. The authors state that, at least close to  threshold, the 
dominant contribution to the hyperon production is the $N^*(1650)$ resonance. 

	Gasparian et al.~\cite{gas00} performed calculations within the 
J\"ulich Meson Exchange Model, where $\pi$ and $K$-exchange was assumed 
including the interference between these two amplitudes (dashed line). It is 
observed by the authors, that in the case of the $\Lambda$ production 
$K$-exchange is dominant and consequently constructive or destructive 
interference between $\pi$ and $K$-exchange give similar results. For the 
$\Sigma^0$ production, however, the 
strength of the contributions from $\pi$ and $K$ exchange are comparable 
resulting in a strong reduction of the $\Sigma^0$ production with a 
destructive interference by which the observed cross sections at threshold are
reproduced. Within their calculations~\cite{gas00,gasp:2000jh} the energy 
dependence of the cross section for other isospin channels are predicted like 
the $\Sigma^+$ production in the reaction $pp \to n K^+ \Sigma^+$. Here, the 
predicted behavior of the cross sections for destructive and constructive 
$\pi$ and $K$-exchange is opposite to that observed in $\Sigma^0$ production.
For a destructive interference the cross section for 
$pp \rightarrow n K^+\Sigma^+$ is expected to be a factor of three higher and 
for constructive interference a factor of three lower than the cross section 
for $pp \rightarrow p K^+\Sigma^0$.

	Data in the other isospin channels will help to extract
the dominant mechanisms in the threshold hyperon production. 

	Measurements of the $pp \to n K^+ \Sigma^+$ reaction have been already 
performed at the COSY--11 facility. The data are presently under 
analysis~\cite{c11:prop}.

\subsection{Effective range parameters}

	The final state interactions of a two body subsystem in a 3-body final
state like $pK^+Y$ influence the excitation function in the threshold region
and its analysis allows to extract information on the effective range 
parameters (for review see ref.~\cite{revart}). A parametrisation of the cross 
section which relates the shape of the threshold behavior to the effective 
range parameters is e.g. \ given by F\"aldt and Wilkin~\cite{fael97}:
\vspace{-0.4cm}        

\be\label{faldtwilkinflux}
\sigma \;=\; 
const\cdot \frac{V_{ps}}{\mbox{F}} \cdot 
  \frac{1}
    {\left(1\;+\;\sqrt{1\,+\,\frac{\mbox{{\tiny Q}}}{\epsilon^\prime}}\right)^2} 
=\;C^{\prime} \cdot 
  \frac{\mbox{Q}^{2}}{ \sqrt{\lambda(\mbox{s},\mbox{m}_p^2,\mbox{m}_p^2)}} 
  \cdot 
  \frac{1}
    {\left(1\;+\;\sqrt{1\,+\,\frac{\mbox{\tiny Q}}{\epsilon^\prime}}\right)^2}\;.
\end{equation}
The phase space volume $V_{ps}$ and the flux factor F are given by~\cite{byckling}:
\begin{equation}\label{Vps_nonrelativistic}
V_{ps} = \frac{\pi^{3}}{2} 
  \frac{\sqrt{\mbox{m}_{p}\,\mbox{m}_{K^+}\,\mbox{m}_{Y}}}
       {(\mbox{m}_{p} + \mbox{m}_{K^+} + \mbox{m}_{Y})^\frac{3}{2}} \; \mbox{Q}^2,\hspace{1cm}
\mbox{F} = 
  2\,(2\pi)^{3n-4}\;\sqrt{\lambda(\mbox{s},\mbox{m}_p^2,\mbox{m}_p^2)}\;.
\end{equation}
with the triangle function 
$\lambda(x, y, z) = x^2 + y^2 + z^2 - 2xy - 2yz - 2zx$.

The results of $\chi^2$ fits using the F\"aldt and Wilkin formula are 
presented in figure~\ref{cros:plot}a) by the solid lines (the dotted lines 
correspond to pure S-wave phase space distributions).
The parameter 
$\epsilon^{\prime}$, which is related to the strength of the $p-Y$ final state
interaction, and the normalization constant $C^{\prime}$ were extracted by the
fits performed for each reaction separately resulting in:
\vspace{0.2cm}

\hspace{2cm}$C^{\prime}( \Lambda)$    =        (98.2~$\pm$~3.7) nb/MeV$^2$ 
\hspace{2cm}$\epsilon^{\prime} (\Lambda)$  = ($5.51^{\,+0.58}_{\,-0.52}$) MeV 
\vspace{0.15cm}

\hspace{2cm}$C^{\prime}(\Sigma^0)$   =        (2.97~$\pm$~0.27) nb/MeV$^2$ 
\hspace{1.6cm}$\epsilon^{\prime} (\Sigma^0)$ = ($133^{\,+108}_{\,-44}$) MeV. 
\vspace{0.2cm}

	Assuming only S-wave production, the $p-\Lambda (\Sigma^0)$ systems 
can be described using the Bergman potentials~\cite{newt82}, where scattering 
length~$\hat a$ and effective range~$\hat r$ are given by:
\begin{equation}\label{scat_par_colin}
  \hat a = \frac{\alpha + \beta}{\alpha \beta}\:, \quad \quad
  \hat  r = \frac{2}{\alpha + \beta}\:,
\end{equation}
with a shape parameter $\beta$, and  $\epsilon^{\prime}~=~\alpha^2/ 2 \mu$ 
where $\mu$ is the reduced mass of the $p-Y$ system~\cite{newt82}. The 
negative value of $\alpha$ is chosen since (at least for $p-\Lambda$) an 
attractive interaction is expected~\cite{hol89,Rij:1998}.

\begin{figure}[h]
\epsfig{figure=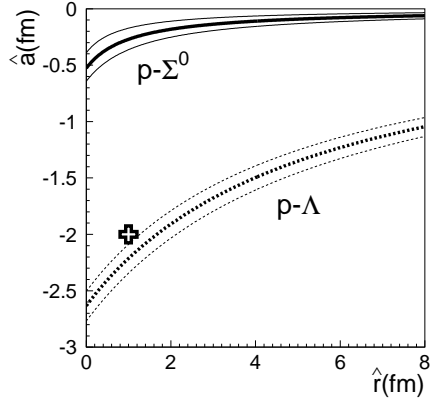,scale=0.3}
\caption{Correlation between the $p-\Sigma^0$  (solid lines) 
and $p-\Lambda$ (dashed lines) effective range parameters.}
\label{a:od:r:plot}
\end{figure}

The parameters $\hat a$ and $\hat r$ are interdependent and only
correlations between them can be 
deduced. In figure~\ref{a:od:r:plot} the correlations obtained for the 
$p-\Sigma^0$ and $p-\Lambda$ systems are presented  by solid and dashed lines, 
respectively. The errors in $\epsilon^{\prime}$ are reflected in the error 
ranges and shown in the figure by the thinner lines. The cross symbol 
represents the singlet and triplet averaged value of the $p-\Lambda$ 
scattering length and effective range parameters extracted from a FSI approach
in threshold $\Lambda$ production~\cite{bal98b}. 

	It seems that the $p-\Sigma^0$ FSI are much smaller than the FSI
for $p-\Lambda$ system.

\vspace{-0.3cm}
\vspace{-.7cm}

\section{Summary}

	Measurements of the energy dependence of the total cross sections for 
the \pppKL and \pppKSo production performed at the COSY--11 facility at excess
energies between 14 and 60 MeV show that the cross section ratio strongly 
decreases in the excess energy range between 10 and 20 MeV.

	Different theoretical models are able to describe the data within a 
factor of two with more or less the same quality, even though they differ in 
the dominant contribution to the production mechanism. Data in the hyperon 
sector are still too limited to put exact constrains on the existing models.

	The new data suggest that the final state interaction in the 
$p-\Sigma^0$ channel is much weaker than in the case of the $p-\Lambda$ 
system.

	Measurements of the hyperon production in other isospin channels like
e.g. the $pp \to nK^+\Sigma^+$ reaction measured recently at COSY--11 will 
help to disentangle the production mechanisms in the threshold region.


\begin{thebibliography}{24}
\expandafter\ifx\csname natexlab\endcsname\relax\def\natexlab#1{#1}\fi
\providecommand{\enquote}[1]{``#1''}
\expandafter\ifx\csname url\endcsname\relax
  \def\url#1{\texttt{#1}}\fi
\expandafter\ifx\csname urlprefix\endcsname\relax\def\urlprefix{URL }\fi

\bibitem[Brauksiepe et~al.(1996)]{bra96}
Brauksiepe, S., et~al., \emph{Nucl. Instr. \& Meth.}, \textbf{A376}, 397--410
  (1996).

\bibitem[Sewerin et~al.(1999)]{sew99}
Sewerin, S., et~al., \emph{Phys. Rev. Lett.}, \textbf{83}, 682--685 (1999).

\bibitem[Baldini et~al.(1988)]{bal88}
Baldini, A., et~al., \emph{Total Cross--Sections for Reactions of High--Energy
  Particles}, Landolt--B\"ornstein, {\em New Series} I/12, Springer, Berlin,
  1988.

\bibitem[Kowina et~al.(2003)]{kow03}
Kowina, P., et~al., \emph{Submited to EPJ A} (2003).

\bibitem[Moskal et~al.(2001)]{pawel:nim}
Moskal, P., et~al., \emph{Nucl. Instr. \& Meth.}, \textbf{A466}, 448--455
  (2001).

\bibitem[Maier(1997)]{maie97}
Maier, R., \emph{Nucl. Instr. \& Meth.}, \textbf{A390}, 1--8 (1997).

\bibitem[Dombrowski et~al.(1997)]{domb97}
Dombrowski, H., et~al., \emph{Nucl. Instr. \& Meth.}, \textbf{A386}, 228--234
  (1997).

\bibitem[Balewski et~al.(1998{\natexlab{a}})]{bal98a}
Balewski, J., et~al., \emph{Phys. Lett.}, \textbf{B420}, 211--216
  (1998{\natexlab{a}}).

\bibitem[Bilger et~al.(1998)]{tof:blig98}
Bilger, A., et~al., \emph{Phys. Lett.}, \textbf{B420}, 217--224 (1998).

\bibitem[Sibirtsev et~al.(2000)]{sib00b}
Sibirtsev, A., et~al., \emph{e-Print Archive, nucl-th/0004022} (2000).

\bibitem[Tsushima et~al.(1999)]{tsu99}
Tsushima, K., Sibirtsev, A., and Thomas, A.~W., \emph{Phys. Rev.},
  \textbf{C59}, 369--387 (1999), erratum-ibid.C61:029903,2000.

\bibitem[Sibirtsev(1995)]{sib95}
Sibirtsev, A., \emph{Phys. Lett.}, \textbf{B359}, 29--32 (1995).

\bibitem[Tsushima et~al.(1997)]{tsu97}
Tsushima, K., Sibirtsev, A., and Thomas, A.~W., \emph{Phys. Lett.},
  \textbf{B390}, 29--35 (1997).

\bibitem[Shyam et~al.(2001)]{shy01}
Shyam, R., Penner, G., and Mosel, U., \emph{Phys. Rev.}, \textbf{C63}, 022202
  (2001).

\bibitem[Gasparian et~al.(2000)]{gas00}
Gasparian, A., et~al., \emph{Phys. Lett.}, \textbf{B480}, 273--279 (2000).

\bibitem[Gasparian et~al.(2001)]{gasp:2000jh}
Gasparian, A., et~al., \emph{Nucl. Phys.}, \textbf{A684}, 397--399 (2001).

\bibitem[Ro\.zek and Grzonka(2002)]{c11:prop}
Ro\.zek, T., and Grzonka, D., \emph{COSY Proposal}, \textbf{117} (2002).

\bibitem[Moskal et~al.(2002)]{revart}
Moskal, P., Wolke, M., Khoukaz, A., and Oelert, W., \emph{Prog. Part. Nucl.
  Phys.}, \textbf{49}, 1--90 (2002).

\bibitem[F\"aldt and Wilkin(1997)]{fael97}
F\"aldt, G., and Wilkin, C., \emph{Z. Phys.}, \textbf{A357}, 241--243 (1997).

\bibitem[Byckling and Kajantie(1973)]{byckling}
Byckling, E., and Kajantie, K., \emph{{Particle Kinematics}}, John Wiley $\&$
  Sons Ltd., 1973, iSBN~0~471~12885~6.

\bibitem[Newton(1982)]{newt82}
Newton, R.~G., \emph{Scattering Theory of Waves and Particles},
  Springer-Verlag, New York, 1982.

\bibitem[Holzenkamp et~al.(1989)]{hol89}
Holzenkamp, B., Holinde, K., and Speth, J., \emph{Nucl. Phys.}, \textbf{A500},
  485--528 (1989).

\bibitem[Rijken et~al.(1999)]{Rij:1998}
Rijken, T.~A., Stoks, V. G.~J., and Yamamoto, Y., \emph{Phys. Rev.},
  \textbf{C59}, 21--40 (1999).

\bibitem[Balewski et~al.(1998{\natexlab{b}})]{bal98b}
Balewski, J., et~al., \emph{Eur. Phys. J.}, \textbf{A2}, 99--104
  (1998{\natexlab{b}}).

\end{thebibliography}
\end{document}